# Shift of Infrared Absorption and Emission Spectra of Transition Metal Ions in Solid Solutions of Semiconductor Compounds

S. V. Naydenov

*Institute of Single Crystals, National Academy of Sciences of Ukraine, Kharkov 61001, Ukraine*
*e-mail: sergei.naydenov@gmail.com*

**Abstract**

A universal theoretical model is proposed that explains the observed shift of IR absorption and emission bands in the spectra of transition metal ions in solid solutions of semiconductor compounds. The model has been used for estimating the long-wavelength shift of luminescence bands in the spectra of semiconductor solid solutions with increasing concentration in application to crystals of the ternary systems $Zn_{1-x}Mg_xSe:Cr^{2+}$ and $Cd_{1-x}Mn_xTe:Fe^{2+}$. Description of the phenomenon is generalized to the case of multicomponent solid solutions.

Crystalline materials based on semiconductor compounds of Groups II and VI doped with transition metal ions ($A^{II}B^{VI}:TM^{2+}$) belong to promising laser media for generation in the near- and mid-IR range [1, 2]. Various applications require developing IR laser media with broad tunable generation bands shifted toward long wavelengths (by up to 5–6 μm). New possibilities in comparison with binary compounds are presented by solid solutions of complex chalcogenides in which the main cation in crystal lattice is substituted with a lighter atom. In particular, some interesting variants are offered by the crystals of $Zn_{1-x}Mg_xSe:Cr_{2+}$ [3], $Zn_{1-x}Mn_xS:Fe_{2+}$ [4], $Cd_{1-x}Mn_xTe:Cr_{2+}$ [5], and $Cd_{1-x}Mn_xTe:Fe_{2+}$ [6, 7], in which lasing in the mid-IR range has been obtained. A distinctive feature of these crystals consists in a significant shift of the absorption and luminescence bands of transition metal ions depending on the cation concentration in solid solution (~20–50 nm per every 10% increment of cation content $x$) and (to a lower degree) on the temperature. These effects open the possibility of control over the parameters of lasing and the shift of generation band toward longer wavelengths. Since the physical nature of this phenomenon is not yet completely clear, the present work was aimed at constructing a universal theoretical model so as to explain the observed effects and relate microscopic parameters of the electron energy band structure to experimental data.

Let us consider the luminescence of isovalent transition metal impurities $TM^{2+}$ (TM = Cr, Fe, Co, Ni, etc.) used to dope crystals of $A^{II}B^{VI}:TM^{2+}$ semiconductor compounds. Since the differences between ion radii and electronegativities of $TM^{2+}$ and crystal-lattice cations (Zn, Cd, Hg, etc.) usually do not exceed 10–15% and 0.4–0.6 Poling units, respectively, the $TM^{2+}$ ion can, according to the Goldschmidt rule, isomorphous substitute a crystal-lattice cation. As a result, the "central" ion occurs in the octahedral or tetrahedral (depending on the lattice type) environment of neighboring chalcogen (S, Se, Te) anions. The energy levels of this impurity ion fall deep into the semiconductor bandgap [8, 9]. The luminescence of $TM^{2+}$ ions in the IR range has the intra-center character. Note that processes of charge exchange between $TM^{2+}$ ions under irradiation with visible and UV light or due to impact ionization by free electrons in a strong (several kilovolts) electric field and related recombination (short-wavelength) luminescence will not be considered below.

The crystalline field of ligands produces splitting of energy levels of the central ion (see, e.g., [10]). For an isolated tetrahedral complex, splitting energy $\Delta E$ can be expressed by the Bethe formula (in SI units):

$$\Delta E^* = \frac{20}{27} \frac{e^2 Z^2}{4\pi\varepsilon_0 \varepsilon} \frac{\langle r^4 \rangle}{d^{*5}}, \qquad (1)$$

where $\varepsilon_0 = 8.85 \times 10^{-12}$ F/m is the dielectric constant of vacuum, $\varepsilon$ is the dielectric constant of the medium, $e = 1.6 \times 10^{-19}$ C is the electron charge, $Z$ is the ligand charge (equal to the ion charge) in $e$ units, $\langle r^4 \rangle$ is the matrix element of $r^4$ ($r$ being the distance of valence $d$-electron from the ion nucleus) calculated from the wave function of the many-body system involving the central ion with all electrons, and $d^*$ is the distance between central ion and ligands (i.e., bond length in the crystal lattice). For the cubic lattice of a binary (sphalerite type) compound, $d^* = (\sqrt{3}/4)\ a$. In the general case, it can be assumed that $d^*$ is proportional to the crystal lattice period $d$.

**Table 1.** Experimental parameters [2] of $A^{II}B^{VI}$:$TM^{2+}$ binary crystals (TM = Cr, Fe) and the corresponding calculated values of splitting energies $\Delta E$ for intra-center transitions in $TM^{2+}$ ions

| Parameter | Doped with $Cr^{2+}$ | | | | | Doped with $Fe^{2+}$ | | | |
|---|---|---|---|---|---|---|---|---|---|
| | ZnS | ZnSe | ZnTe | CdS | CdSe | ZnS | ZnSe | CdSe | CdTe |
| Crystal lattice type | ZB | ZB | ZB | W | W | ZB | ZB | W | ZB |
| Lattice parameter(s), Å | 5.41 | 5.67 | 6.10 | $c$ = 6.75 | $c$ = 7.02 | 5.41 | 5.67 | $c$ = 7.02 | 6.48 |
| | | | | $a$ = 4.14 | $a$ = 4.30 | | | $a$ = 4.30 | |
| Bandgap width, eV | 3.7 | 2.7 | 2.3 | 2.5 | 1.7 | 3.7 | 2.7 | 1.7 | 1.5 |
| Absorption peak $\lambda_{ab}$, μm | 1.69 | 1.77 | 1.79 | 1.85 | 1.92 | 2.8 | 3.1 | 3.5 | 3.65 |
| Emission peak $\lambda_{em}$, μm | 2.35 | 2.45 | 2.4 | 2.6 | 2.75 | 3.94 | 4.35 | 4.81 | 5.94 |
| Energy of splitting $\Delta E$, eV | 0.464 | 0.442 | 0.438 | 0.393 | 0.371 | 0.286 | 0.253 | 0.216 | 0.192 |

\* ZB = zinc blende type; W = wurtzite type.

For $A^{II}B^{VI}$:$TM^{2+}$ binary crystals, formula (1) can be considered as a good approximation for estimating the energy gap between electron levels in the ground and excited states of a $TM^{2+}$ ion. The spin–orbit interaction in distorted lattice and the interaction of lattice with electron subsystem (vibronic excitations) lead due to the Jahn–Teller effect to an additional splitting of $d$-electron levels in $TM^{2+}$ ion, which smear into rather broad sub-bands. Usually, the splitting energy is $\Delta E \propto 10^3$ cm$^{-1}$, while the splitting of vibronic levels is $\Delta E_v \propto$ 1–10 cm$^{-1}$.

The intra-center transition in $TM^{2+}$ ion during the absorption and emission in the IR range takes place between sub-bands of the ground and excited states of ion separated by energy gap $\Delta E$. Formula (1) indicates that $\Delta E$ decreases with increasing $d$ value and, hence, with growing lattice parameter of $A^{II}B^{VI}$:$TM^{2+}$ binary compound. Therefore, maxima of the absorption and emission bands must shift toward longer wavelengths. Table 1 presents the experimental and calculated data (computation details omitted) for a series of binary laser crystals (ZnS, ZnSe, ZnTe, CdS, CdSe, CdTe) doped with chromium and iron ions. Upon passage from light to heavy cations and/or anions, covalent bonds in the lattice are weakening (replaced by ionic), the lattice parameter grows (and the bandgap width drops since covalent bonds are weakening), the energy gap $\Delta E$ of the intra-center transition decreases, and the IR absorption and emission bands consistently shift toward longer wavelengths. Despite the simplicity and obviousness of the latter conclusion (cited in most works devoted to $A^{II}B^{VI}$:$TM^{2+}$ laser crystals), this regularity is completely absent in crystals of the solid solutions of ternary compounds. For example, the lattice constant in $Zn_{1-x}Mg_xSe$:$Cr^{2+}$ crystals grows with increasing concentration of magnesium; on the contrary, the lattice constant in $Cd_{1-x}Mn_xTe$:$Fe^{2+}$ crystals drops with increasing content of manganese. At the same time, both crystals exhibit long-wavelength shift of the IR absorption and emission bands [3–7].

In substitutional solid solutions of the $A_{1-x}B_xC$:$TM^{2+}$ type (where A and B are cations of the host lattice and diluent, respectively; C is a chalcogen; and $x$ is the solid solution concentration), the impurity ion can replace any of the two kinds of cations (rather than a single one). Since splitting energy $\Delta E$ of $TM^{2+}$ ion levels depends on the local environment, it cannot be expressed by formula (1). To relate this energy value to parameters of the semiconductor material, let us naturally assume that, on the passage from simple compounds to their solid solution, a physical quantity should vary (to the first approximation) in proportion to the concentrations of components in the mixed system. In a particular case, this principle as applied to lattice parameter $d(x)$ corresponds to the well-known Vegard law $d(x) = d_1 + (d_2 - d_1)x$, which is quite valid for the materials under consideration.

In application to energy gap $\Delta E$ between states of the $TM^{2+}$ ion, this principle of additivity implies that

$$\Delta E_{ABC:TM}(x) = (1-x)\Delta E_{AC:TM} + x\Delta E_{BC:TM}. \quad (2)$$

Note that, in the framework of this approach, it is also possible to take into account the effect of temperature by adding the term $\propto T$ (not considered here). In different notation, Eq. (2) can be rewritten as follows:

$$\varepsilon(x) = \varepsilon_1 - \delta_{12}x; \quad \delta_{12} = \varepsilon_1 - \varepsilon_2, \quad (3)$$

where energy $\varepsilon(x)$ corresponds to the energy of photon absorbed or emitted during the intra-center transition in $TM^{2+}$ ion in the matrix of solid solution with concentration $x$.

**Table 2.** Parameters of binary components ($i = 1, 2$) of ABC:$TM^{2+}$ solid solutions (dielectric permittivity $\varepsilon$, ion radius $r$ of cation, lattice parameter $d$) and parameters of long-wavelength shift ($\delta_{12}$ and $\kappa$) for intra-center transitions in $TM^{2+}$ ions (radius $r_1$ of a valence $d$-electron in a TM ion approximated by impurity atom radius)

| Crystal | $\varepsilon_1$ | $\varepsilon_2$ | $r_1$, Å | $r_2$, Å | $d_1$, Å | $d_2$, Å | $r_l$, Å | $\delta_{12}$, eV | $\kappa$ |
|---|---|---|---|---|---|---|---|---|---|
| ZnMgSe:$Cr^{2+}$ | 7.32 | 4.66 | 0.74 | 0.71 | 5.67 | 6.88 | 1.27 | $6.84 \times 10^{-2}$ | $1.55 \times 10^{-1}$ |
| CdMnTe:$Fe^{2+}$ | 12.82 | 19.30 | 0.92 | 0.80 | 6.48 | 6.33 | 1.26 | $1.22 \times 10^{-2}$ | $6.36 \times 10^{-2}$ |

Wavelength shift $\Delta\varepsilon$ is related to the photon energy change as

$$\Delta\lambda/\lambda = -\Delta\varepsilon/\varepsilon. \quad (4)$$

In the case of small increments, $\Delta\lambda \ll \lambda$ or $\Delta\varepsilon \ll \varepsilon$, it follows from Eqs. (3) and (4) that

$$\Delta\lambda \approx \lambda\kappa\Delta x; \quad \kappa = \delta_{12}/\varepsilon \ll 1, \quad (5)$$

which is the law of linear shift of the maximum of absorption or emission band depending on change $\Delta x$ in the concentration of solid solution. Here, $\lambda$ values correspond to maxima in the spectra of absorption ($\lambda_{ab}$) or emission ($\lambda_{em}$) of the "undiluted" binary compound. The character of the spectral shift is determined by the sign of $\delta_{12}$, whereby the long wavelength and short wavelength correspond to $\delta_{12} > 0$ and $\delta_{12} < 0$, respectively. In the general case, the value of shift coefficient $\kappa = \kappa_\alpha$ depends on the structure of energy band spectrum of a particular TM ion in the matrix of solid solution. For each separate $\alpha_{th}$ band in a complex spectrum of absorption or emission, this coefficient can take different (albeit usually close) values. The main contribution is related to the Stark splitting — that is, to the $\Delta E(x)$ value.

Table 2 presents simplified estimates of parameter $\delta_{12}$ calculated for $Zn_{1-x}Mg_xSe:Cr^{2+}$ and $Cd_{1-x}Mn_xTe:Fe^{2+}$ ternary crystals. Both compounds have $\delta_{12} > 0$, so that the bands shift toward longer wavelength. As can be seen from Tables 1 and 2, the orders of magnitude for band-shift parameters are $\varepsilon \propto 10^{-1}$ eV, $\delta \propto 10^{-2}$ eV, and $\kappa \propto 10^{-1}$.

The IR range corresponds to wavelengths $\lambda \propto 10^3$ nm, from which it follows that the maximum bandshift due to concentration changes on the order of $\Delta x \propto 1$ can reach several hundred nanometers ($\Delta\lambda \propto 10^2$ nm). This estimation is also valid (in the order or magnitude) for other IR laser crystals based on ternary chalcogenides. The "red" shift of luminescence band maxima in $Cd_{1-x}Mn_xTe:Fe^{2+}$ ternary crystals is several times smaller than that in $Zn_{1-x}Mg_xSe:Cr^{2+}$ crystals because of their strongly different $\delta_{12}$ values and 2.4-times lower shift coefficient $\kappa$ (in the framework of theoretical estimation), which agrees with the experimentally observed 2- to 2.5-times decrease in the band shift.

Linear relation (5) and related estimates have been confirmed by experimental data (see, e.g., [3–7]), which established that there are long-wavelength shifts of the absorption and luminescence bands by $\propto$ 50 nm (for $Zn_{1-x}Mg_xSe:Cr^{2+}$ crystals) and $\propto$ 20–25 nm (for $Cd_{1-x}Mn_xTe:Fe^{2+}$ crystals) per every 10% change in the concentration of solid solution (of Mg or Mn, respectively). From this it follows that the proportionality coefficients for the linear shift of absorption and emission bands are $\kappa_{ab} \approx 2.82 \times 10^{-1}$, $\kappa_{em} \approx 2.04 \times 10^{-1}$ for $Zn_{1-x}Mg_xSe:Cr^{2+}$ (absorption and emission peaks at 1770 and 2450 nm, respectively) and $\kappa_{ab} \approx 6.85 \times 10^{-2}$, $\kappa_{em} \approx 4.21 \times 10^{-2}$ for $Cd_{1-x}Mn_xTe:Fe^{2+}$ (absorption and emission peaks at 3650 and 5490 nm, respectively) in agreement with the calculated data presented in Table 2. Differences are related to the rough character of estimation of the parameters of crystal field, which took into account only the Stark splitting and ignored further splitting of degenerate states in the ground and excited zone of TM ion energy levels. This neglect leads to smearing (broadening) of the spectrum of TM ion levels and splitting of the absorption and emission spectra into several overlapping bands. The parameters of this modification of the energy band structure depend on the concentration of solid solution. However, linear dependence (5) with corresponding shift coefficient $\kappa_\alpha$ is valid for each particular band.

It should be noted that impurities and defects present in solid solutions of $A^{II}B^{VI}$:$TM^{2+}$ compounds can influence the properties of the energy-band structure and characteristics of luminescence, especially in the short-wavelength region. The conditions of laser crystal operation in the IR spectral range require a high degree of their purification from foreign impurities capable of increasing the concentration of free carriers and impairing the IR-transparency of crystals as a result of light scattering on electrons. In high-quality laser crystals based on $A^{II}B^{VI}$ solid solutions, the concentration of undesired impurities is usually below $10^{-2}$–$10^{-5}$% (3N to 6N degree of raw material purity), which is significantly smaller than the concentration of active TM ions in the matrix (which is on a level of several percent). In such crystals, the influence of foreign impurities on the phenomena under consideration can be ignored.

The law established above can be expanded so as to apply to multicomponent solid solutions of the type $A(x_1)\{B_i(x_i)\}C(y_1)\{D_j(y_j)\}$:$TM^{2+}$, which contain, in addition to the main dopant cation with concentration $x_1$ and main anion with concentration $y_1$, additionally $i = 2, ..., N$ cations with concentrations $x_i$ and $j = 2, ..., M$ anions with concentrations $y_j$. For example, these systems include the crystals of $Zn_{1-x}Mg_xSe_{1-y}S_y$:$Cr^{2+}$ and $Cd_{1-x}Mn_xTe_{1-y}Se_y$:$Fe^{2+}$. The expected shift of energy gap between the $TM^{2+}$ ion states is

$$\varepsilon(\{x_i\},\{y_j\}) = \varepsilon_{11} - \sum_{i=2}^{N}\delta_i x_i - \sum_{j=2}^{M}\gamma_j y_j + \sum_{i=2}^{N}\sum_{j=2}^{M}\xi_{ij} x_i y_j. \quad (6)$$

where $\varepsilon_{11}$ is splitting energy $\Delta E$ for the main component ac and the other parameters are defined as

$$\delta_i = \delta_i(y_1) = \varepsilon_{11} - y_1 \varepsilon_{i1}; \quad \gamma_j = \gamma_j(x_1) = \varepsilon_{11} - x_1 \varepsilon_{1j};$$

$$\xi_{ij} = \varepsilon_{11} + \varepsilon_{ij} \geq 0. \quad (7)$$

Here, $\varepsilon_{i1}$, $\varepsilon_{1j}$, $\varepsilon_{ij}$ are the values of $\Delta E$ for binary components $B_iC$, $AD_j$, and $B_iD_j$, respectively. The concentration of solid solution components obey the normalization condition

$$x_1 = 1 - \sum_{i=2}^{N} x_i, \quad y_1 = 1 - \sum_{j=2}^{M} y_j. \quad (8)$$

In a multicomponent system with almost "equitable" partial composition, we have $x_i \propto 1/N$ and $y_j \propto 1/M$, so that the nonlinear term $\propto x_i y_j \ll 1$ in Eq. (6) can be ignored for $N > 2$ and $M > 2$. The expected "quasi-linear" shift of the absorption and emission bands can then be expressed as

$$\Delta\lambda(\{x_i\},\{y_j\}) \approx \lambda\left[\sum_{i=2}^{N}\kappa_i \Delta x_i + \sum_{j=2}^{M}\nu_j \Delta y_j\right];$$

$$\kappa_i = \frac{\delta_i}{\varepsilon_{11}}, \quad \nu_j = \frac{\gamma_j}{\varepsilon_{11}}. \quad (9)$$

Depending on the signs of parameters $\delta_i$ and $\gamma_j$, there are $N + M - 2$ types of behavior of the IR spectra of solid solutions with variable concentrations of components. In the simplest case of $N = 2$, $M = 1$, formula (9) converts into law (5) for a ternary compound. The maximum shift of IR spectra corresponds to a compound for which shift parameters $\delta_i$ and $\gamma_j$ (and, hence, $\kappa_i$ and $\nu_j$) take the maximum values:

$$\Delta\lambda(\{x_i\},\{y_j\}) \leq \lambda[\kappa_{max}\Delta x_{max} + \nu_{max}\Delta y_{max}]. \quad (10)$$

The spectral shift for ternary compounds can also be experimentally estimated based on the measurements of peaks in the spectra of absorption ($\lambda_{ab}$) or emission ($\lambda_{em}$) for two samples of crystals with different concentrations of solid solution ($x_1 \neq x_2$), using the formula

$$\Delta\lambda = \frac{\lambda_2 - \lambda_1}{x_2 - x_1}\Delta x, \quad (11)$$

where one concentration may correspond to the binary composition ($x_1 = 0$). In ($N + M$)-component system, this approach will require $N + M - 2$ measurements for restoring all band-shift parameters in Eq. (9), i.e., for solving the "inverse" problem.

The established relations can be used for predicting the expected shifts of IR luminescence bands for solid solutions of chalcogenides without growing these crystals. This can be useful in the development of new laser media for the mid-IR spectral range.

REFERENCES

1. I. T. Sorokina, in *Solid-State Mid-Infrared Laser Sources*, Ed. by I. T. Sorokina and K. L. Vodopyanov, Top. Appl. Phys. **89**, 255 (2003). https://doi.org/10.1007/3-540-36491-9_7
2. S. B. Mirov, I. S. Moskalev, S. Vasilyev, V. Smolski, V. V. Fedorov, D. Martyshkin, J. Peppers, M. Mirov, A. Dergachev, and V. Gapontsev, IEEE J. Sel. Top. Quantum Electron. **24**, 1601829 (2018). https://doi.org/10.1109/JSTQE.2018.2808284
3. M. E. Doroshenko, V. V. Osiko, H. Jelinkova, M. Jelinek, M. Nemec, J. Sulc, N. O. Kovalenko, A. S. Gerasimenko, and V. M. Puzikov, Opt. Mater. **47**, 185 (2015). https://doi.org/10.1016/j.optmat.2015.05.015
4. A. D. Martinez, D. V. Martyshkin, R. P. Camata, V. V. Fedorov, and S. B. Mirov, Opt. Mater. Express **5**, 2036 (2015). https://doi.org/10.1364/OME.5.002036
5. U. Hommerich, X. Wu, V. R. Davis, S. B. Trivedi, K. Grasza, R. J. Chen, and S. Kutcher, Opt. Lett. **22**, 1180 (1997). https://doi.org/10.1364/OL.22.001180
6. M. E. Doroshenko, V. V. Osiko, H. Jelinkova, M. Jelinek, J. Sulc, D. Vyhlidal, N. O. Kovalenko, and I. S. Terzin, Opt. Mater. Express **8**, 1708 (2018). https://doi.org/10.1364/OME.8.001708
7. M. E. Doroshenko, H. Jelinkova, M. Jelinek, J. Sulc, D. Vyhlidal, N. O. Kovalenko, and I. S. Terzin, Opt. Lett. **43**, 5058 (2018). https://doi.org/10.1364/OL.43.005058
8. A. Zunger, in Solid State Physics, Ed. by H. Ehrenreich and D. Turnbull (Academic Press, 1986), **Vol. 39**, p. 275. https://doi.org/10.1016/S0081-1947(08)60371-9
9. T. P. Surkova, M. Godlewski, K. Swiatek, P. Kaczor, A. Polimeni, L. Eaves, and W. Giriat, Phys. B (Amsterdam, Neth.) **273–274**, 848 (1999). https://doi.org/10.1016/S0921-4526(99)00519-0
10. I. B. Bersuker, *Electronic Structure and Properties of Coordination Compounds. Introduction to Theory*, 3rd ed. (Khimiya, Leningrad, 1986), p. 69 [in Russian].